\documentclass[12pt]{iopart}
\usepackage{iopams}
\usepackage{graphics}
\usepackage{graphicx}
\usepackage{epsfig}
\usepackage{dcolumn}
\usepackage{bm}
\begin{document}

\title[Enhanced confinement with increased extent of the low magnetic shear region]{Enhanced confinement with increased extent of the low magnetic shear region in tokamak plasmas}

\author{Lorenzo Nasi, Marie-Christine Firpo}

\address{Laboratoire de Physique et Technologie des Plasmas
(CNRS UMR 7648), Ecole Polytechnique, 91128 Palaiseau cedex, France}

\ead{firpo@lptp.polytechnique.fr}
\begin{abstract}
The Hamiltonian representation of magnetic field lines enables to
study their confinement properties in tokamaks through the use of
symplectic maps such as the symmetric tokamap and its bounded
version, the latter being introduced here. In this time-independent
purely magnetic framework, we observed the drastic improvement on
the confinement of magnetic field lines produced by the local
vanishing of the shear profile. This amounts to a non-twist
condition that notably acts in the same way the safety profile being
(non-strictly) monotonic or having reversed-shear. We single out the
effect of the amount of flatness of the safety profile in the
vicinity of its zero shear point. All other things being equal, the
beneficial effect of the vanishing of the shear profile is shown to
be increased if the radial extent of the low-shear region is
increased. To be specific, the low-shear region induces the
formation of a belt of robust KAM tori acting as an internal
transport barrier whose width is all the larger as the extent of the
low-shear region is broad.
\end{abstract}

\pacs{52.55.Fa,05.45.+a,52.25.Gj} \submitto{Plasma Physics and
Controlled Fusion} \maketitle

\section{Introduction}

In order to approach ignition, it is required that the present hot
tokamak plasmas have a better energy confinement. The energy
confinement time is commonly limited by deleterious small-scale
turbulent transport. Because internal transport barriers (ITB)
\cite{WolfRevue,ConnorReviewITB2004,Tala06} reduce or even quench
the turbulent transport within the inner plasma region improving
energy confinement in the plasma core which should yield high fusion
gains, they have become an essential ingredient of advanced
scenarios in which the tokamak could operate as an economic
steady-state fusion reactor.

The physics of the triggering and sustainment of ITB as well as the
explanation of their very nature are still subjects of intensive
investigations \cite{Tala06}. Besides an increasingly detailed
empirical and experimental knowledge, their theoretical modeling is
still incomplete. It is commonly admitted that two ingredients
should control their formation, namely the $\mathbf{E}\times
\mathbf{B}$ flow shear and the magnetic shear through the safety
factor ($q$) profile. An increased number of works have recently
reported the importance of the $q$-profile. In particular, Eriksson
et al. \cite{Eriksson02} demonstrated experimentally in the JET
tokamak the critical role of the safety factor profile in the
formation of ITB by using it as a control parameter in discharges
where other parameters, including the $\mathbf{E}\times \mathbf{B}$
flows, were kept the same. They obtained an ITB for a non-monotonic
$q$ profile (i.e. with reversed shear) but not with a monotone $q$.
Actually these reversed shear scenarios have been recognized as
favorable experimental conditions to obtain ITB whereas the ITB's
locations have been mostly related to the low shear zones.

In this article, we wish to study the appearance and robustness of
ITB within a purely magnetic and static (time-independent) framework
following a line of approach notably initiated by Balescu and
coworkers. This simplified, yet relevant, approach is guided by the
fact that, in a first approximation, charged particles in a tokamak
follow the magnetic field lines so that transport properties for the
laters transfer to the formers. We shall focus on the influence of
the behaviour of the $q$-profile in the vicinity of its vanishing
shear region on the appearance and robustness of ITB. To be more
specific, we wish to single out the role and importance of the
amount of \emph{flatness} of the $q$-profile in the vicinity of its
minimum (for the reversed shear case) or in the vicinity of its
inflexion point (in the case of a non-strictly monotonic profile).

In Section \ref{sec-Hamiltonian_maps}, the Hamiltonian formulation
of magnetic field lines is recalled and symplectic maps are
introduced. In order to capture more accurately the magnetic field
lines, we shall implement the symmetric tokamap and its bounded
version, instead of the standard tokamap. In Section
\ref{results_poly}, we shall study numerically the influence of the
flatness of the $q$-profile, in the vicinity of a given inflexion or
minimum point, on the overall magnetic field lines stochasticity. To
do so, we shall consider a family of polynomial winding number
profiles. In Section \ref{sec:realisticProfile}, we eventually
consider a sample of tokamak-oriented realistic $q$-profiles with
reversed shear. This brings another evidence on the favorable role
played by an increased extension of the low-shear region around the
minimum of the $q$-profile. We conclude in Section
\ref{sec:conclusions}.

\section{Hamiltonian formulation for magnetic field lines and
maps} \label{sec-Hamiltonian_maps}

\subsection{Hamiltonian framework}

In order to investigate the role of the $q$-profile on the existence
and robustness of  ITBs, we shall use the Hamiltonian structure of
the magnetic field lines. We shall use in the following traditional
generalized toroidal coordinates $\psi, \theta, \zeta$ where  $\psi$
is the flux coordinate (so that $\psi = 0$ and $\psi=1$ correspond
to the magnetic axis and to the the plasma boundary, respectively),
$\theta$ is the poloidal angle and $\zeta$ the toroidal angle. From
Maxwell's equation $\nabla . \mathbf{B} = 0$, the ``equations of
motion'' for magnetic field lines
\cite{HazeltinePlasmaConfinement2003} in a tokamak are
\begin{equation}  \label{eq:HamiltonTokamak}
 \frac{d\psi}{d\zeta} = - \frac{\partial F}{\partial \theta} \ , \   \frac{d\theta}{d\zeta} =  \frac{\partial F}{\partial \psi}
\end{equation}
where  $F(\psi, \theta, \zeta)$ is the poloidal flux. Equations
(\ref{eq:HamiltonTokamak}) have a Hamiltonian structure: the
toroidal angle $\zeta$ plays the role of ``time'', $F$ is the
Hamiltonian and $\theta$ and $\psi$ are canonically conjugated
coordinates. In order to consider perturbations from the ideal i.e.
integrable case, where $F = F_0(\psi)$ and magnetic surfaces are
nested tori wound around the magnetic axis, let us introduce a
stochasticity parameter, $K$, such that the Hamiltonian $F$ is given
by the sum of an unperturbed term $F_0$ and a perturbation $\delta
F(\psi, \theta, \zeta)$ :
\begin{equation} \label{eq:hamiltionienPerturbe}
    F = F_0(\psi) + K\delta F(\psi, \theta, \zeta ) .
\end{equation}
As $K$ increases, magnetic surfaces may no longer exist and the
problem of their existence is formally equivalent to that of the
integrability of a Hamiltonian with one-and-a-half degrees of
freedom. In realistic tokamaks, these magnetic perturbations may
result e.g. from coil imperfections and/or internal instabilities
(tearing). Within the above formalism, the safety factor $q(\psi)$
is simply related to the unperturbed Hamiltonian since $q(\psi) =
1/W(\psi)$ where $W(\psi) \equiv dF_0/ d\psi$ is the winding number.
Therefore $q(\psi)$,  the ratio of the number of toroidal turns per
poloidal turn, is the inverse of the unperturbed Hamiltonian.
Moreover, $s = d\ln q/d \ln \psi$ is the shear profile.

\subsection{Maps}

\subsubsection{Tokamap}

Instead of performing a time consuming integration of the field line
differential equations (\ref{eq:HamiltonTokamak}), we shall use a
simplified but relevant approach, namely the Poincar\'{e} map
associated to a given poloidal cross-section
\cite{AbdullaevContructionMappings2006}. It is defined from the
intersection points of a magnetic field line starting at position
($\psi_{0}, \theta_{0}$) with some given poloidal section  $\zeta =
cst$. The intersection point after $\nu$ toroidal turns is denoted
by ($\psi_{\nu}, \theta_{\nu}$). The  map we shall be interested in,
the \textit{tokamap}, was introduced by Balescu and coworkers
\cite{BalescuTokamap1998,nobleInternal,WeyssowDemonstrationTokamap1999}.
Its implicit form is
\begin{eqnarray}\label{eq:tokamapImplicit}
 \psi_{\nu + 1} &=& \psi_{\nu } - K\frac{\psi_{\nu + 1}}{1 + \psi_{\nu + 1}} \sin\theta_{\nu} \nonumber \\
\\
\theta_{\nu + 1} &=& \theta_{\nu } + W(\psi_{\nu + 1}) -
\frac{K}{2\pi}\frac{1}{(1 + \psi_{\nu + 1})^2} \cos\theta_{\nu} .
\nonumber
\end{eqnarray}
Explicitly, we have \begin{eqnarray}
 \psi_{\nu + 1} &=& \frac{1}{2}\left( P(\psi_{\nu },\theta_{\nu } ) + \sqrt{[P(\psi_{\nu },\theta_{\nu } )]^{2} + 4 \psi_{\nu }}\right)  \\
 \theta_{\nu + 1} &=& \theta_{\nu } + W(\psi_{\nu + 1}) - \frac{K}{2\pi}\frac{1}{(1 + \psi_{\nu + 1})^2} \cos\theta_{\nu}
\end{eqnarray}
where
\begin{equation}
P(\psi,\theta ) = \psi - 1 -K \sin\theta.
\end{equation}
This map is an area-preserving map which is compatible with the
toroidal geometry \cite{BalescuTokamap1998} due to the following
important physical properties : (i) the toroidal flux $\psi$ is
always positive (this is necessary since $\psi \sim r^2$ where $r$
is the minor radius of the tokamak) and (ii) the magnetic axis is
invariant ($\psi = 0$ is mapped to $\psi = 0$). It has proved to be
an important model since it correctly describes the qualitative
features of the magnetic field lines stochasticity known in tokamak
physics.

\subsubsection{Symmetric tokamap} \label{subsubsymmetric}

Some studies have recently pointed some limitations of this model.
It actually turns out that despite a qualitative agreement with the
actual magnetic field behaviour obtained from the continuous
Hamiltonian system formed by Eqs. (\ref{eq:HamiltonTokamak}) and
(\ref{eq:hamiltionienPerturbe}), the tokamap is far less accurate
than its symmetric version as proposed by Abdullaev
\cite{AbdullaevMappingModels2004,AbdullaevContructionMappings2006}.
We shall therefore implement in this study the symmetric tokamap
instead of the standard one. The symmetric tokamap very closely
describes the continuous Hamiltonian system
(\ref{eq:hamiltionienPerturbe}) with
\begin{equation}
K\delta F(\psi, \theta, \zeta ) =
-\frac{K}{(2\pi)^2}\frac{\psi}{1+\psi} \sum_{n = -M}^{M}\cos(\theta
-n\zeta),
\end{equation}
and  $2M + 1 \gg 1$, from which it is derived. Its implicit form is
given by
\begin{eqnarray}\label{eq:tokamapSymmetricImplicit}
 \Psi_{\nu } &=& \psi_{\nu } - \frac{K}{2\pi}\frac{\Psi_{\nu }}{1 + \psi_{\nu}} \sin\theta_{\nu} \nonumber \\
\Theta_{\nu} &=& \theta_{\nu } - \frac{K}{2\pi}\frac{1}{(1 + \Psi_{\nu})^2} \cos\theta_{\nu}   \nonumber \\
\Theta_{\nu + 1} &=& \Theta_{\nu } + 2\pi W(\Psi_{\nu})\\
\Psi_{\nu+ 1} &=& \Psi_{\nu}\nonumber \\
\psi_{\nu + 1} &=& \Psi_{\nu + 1 } - \frac{K}{2\pi}\frac{\Psi_{\nu +1 }}{1 + \Psi_{\nu +1}} \sin\theta_{\nu +1} \nonumber \\
\theta_{\nu + 1} &=& \Theta_{\nu +1} - \frac{K}{2\pi}\frac{1}{(1 +
\Psi_{\nu+1})^2} \cos\theta_{\nu +1} \nonumber
\end{eqnarray}
where $K$ is the stochasticity parameter and the auxiliary variables
$\Theta$ and $\Psi$  are determined iteratively by the map from a
given starting point $(\theta_{0},\psi_{0})$. Contrarily to the
tokamap, a numerical approximation is necessary in order to solve
the last equation of the symmetric tokamap
(\ref{eq:tokamapSymmetricImplicit}) with respect to variable
$\theta_{\nu + 1}$,  but this is an affordable  task with usual
interpolation methods such as Newton's method or the Brent's
algorithm \cite{BrentAlgo} that we used here.

\subsubsection{Bounded symmetric tokamap}

Another adjustment to the tokamap has also been recently proposed:
the bounded tokamap. Whereas the tokamap does not impose any upper
boundary limit for $\psi$, so that it can be directly used in the
ergodic divertor problem \cite{AbsullaevErgodic99} where magnetic
field lines may go outside the last magnetic surface (defined as
$\psi = 1$), its bounded version is constructed by requiring that
$\psi=1$ cannot be crossed. We used the approach developed in Refs.
\cite{KuzovkovAnnalsCraiova,KuzovkovBoundedTokamap2007} to propose
the symmetric version of the bounded tokamap with
\begin{equation}
 K\delta F(\psi, \theta, \zeta ) = -\frac{K}{(2\pi)^2}\psi(1-\psi) \sum_{n = -M}^{M}\cos(\theta -n\zeta)
\end{equation}
and $2M + 1 \gg1$. This form of perturbation ensures that not only
the magnetic axis, corresponding to $\psi = 0$, but also the plasma
boundary, corresponding to $\psi = 1$ are invariant under the map
and cannot be crossed.

Then, applying the same method
\cite{AbdullaevMappingModels2004,AbdullaevContructionMappings2006}
as in Sec. \ref{subsubsymmetric}, we propose the symmetric version
of the bounded tokamak as
\begin{eqnarray}\label{eq:tokamapSymmetricBoundedImplicit}
 \Psi_{\nu } &=& \psi_{\nu } - \frac{K}{2\pi}\Psi_{\nu }(1 - \Psi_{\nu}) \sin\theta_{\nu} \nonumber \\
\Theta_{\nu} &=& \theta_{\nu } - \frac{K}{2\pi}(1 - 2\Psi_{\nu}) \cos\theta_{\nu}   \nonumber \\
\Theta_{\nu + 1} &=& \Theta_{\nu } + 2\pi W(\Psi_{\nu})\\
\Psi_{\nu+ 1} &=& \Psi_{\nu}\nonumber \\
\psi_{\nu + 1} &=& \Psi_{\nu + 1 } - \frac{K}{2\pi}\Psi_{\nu }(1 - \Psi_{\nu}) \sin\theta_{\nu +1} \nonumber \\
\theta_{\nu + 1} &=& \Theta_{\nu +1} - \frac{K}{2\pi} (1 -
2\Psi_{\nu})\cos\theta_{\nu +1} \nonumber .
\end{eqnarray}
The first equation of (\ref{eq:tokamapSymmetricBoundedImplicit}) has
two roots for  $\Psi_{\nu }$ and the positive one is retained. This
modeling is valid for not too large values of the stochasticity
parameter as it is required that $K/2\pi < 1$.

\subsection{Motivations}

From the theoretical point of view, the effect of various parameters
describing the $q$-profile, such as the boundary values of the
$q$-profile, the value of the minimum of the $q$-profile in the
reversed shear version (in the so-called rev-tokamap), for
Hamiltonian maps have already been considered
\cite{BalescuRevTokamap,ConstantinescuTransportBarriers2005}. Yet,
this has been mainly limited to restricted choices of $q$ profiles.
In addition, to the authors' knowledge, an essential point has not
been investigated systematically: the effect of the local flatness
of the $q$-profile on (hopefully) realistic models for magnetic
field lines in a tokamak. This is in our sense a crucial point since
there is strong experimental evidence that the radial extent of the
low shear region of safety factor's profile plays an outstanding
role on ITBs' appearance and robustness
\cite{Baranov04,ConnorReviewITB2004,Tala06}. Thus, modeling is
needed in order to asses one of the ``\textit{critical physics
issues relevant to the extrapolation of ITB regimes to next-step
experiments, such as ITER} '' \cite{LitaudonITBCriticalIssues2006} :
the optimum $q$-profile. This forms the ultimate perspective of the
present work.

\section{Numerical results on the stochasticity of magnetic field lines for polynomial winding number profiles}\label{results_poly}

We are interested in assessing the influence of the extent of the
low shear region on magnetic confinement. Since it happens that the
Hamiltonian (\ref{eq:hamiltionienPerturbe}) is more directly linked
to the winding number, since $W(\psi) \equiv dF_0/ d\psi$, than the
$q$-profile, we shall consider it in the following. We are then
particularly interested in the influence of the behaviour of the
$W$-profile in the vicinity of some unique $\psi_0$ such that
$W'(\psi_0) = 0$. If such a $\psi_0$ exists, the twist condition for
the corresponding map is violated, that is the symmetric tokamap is
\emph{non-twist}. A straightforward way to study such behaviour is
to consider the family of polynomial $W$-profiles :
 \begin{equation}\label{eq:allureQPolynomiale}
 W(\psi) = -c (\psi - \psi_0)^n + W_{0}
\end{equation}
where  $n$ is a positive integer representing the degree of
degeneracy of the zero $\psi_0$ of the shear profile {\footnote{More
precisely, $\psi_0$ is a zero of the shear profile when $n \geq
2$.}} associated to (\ref{eq:allureQPolynomiale}). Although there is
an extensive bunch of experimental evidence that the maximum value
of the winding number influences ITB formation in the reversed shear
case \cite{Baranov04,ConnorReviewITB2004,Tala06}, this will not be
our point here and we shall rather impose here on an arbitrary basis
$W_{0} = 5/2\pi$, $c = 1.3$ and $\psi_0 = 0.5$ in order to focus on
the flatness of the $W$-profile only.

\begin{figure}[htbp]
    \centering
    \includegraphics[width =  \textwidth]{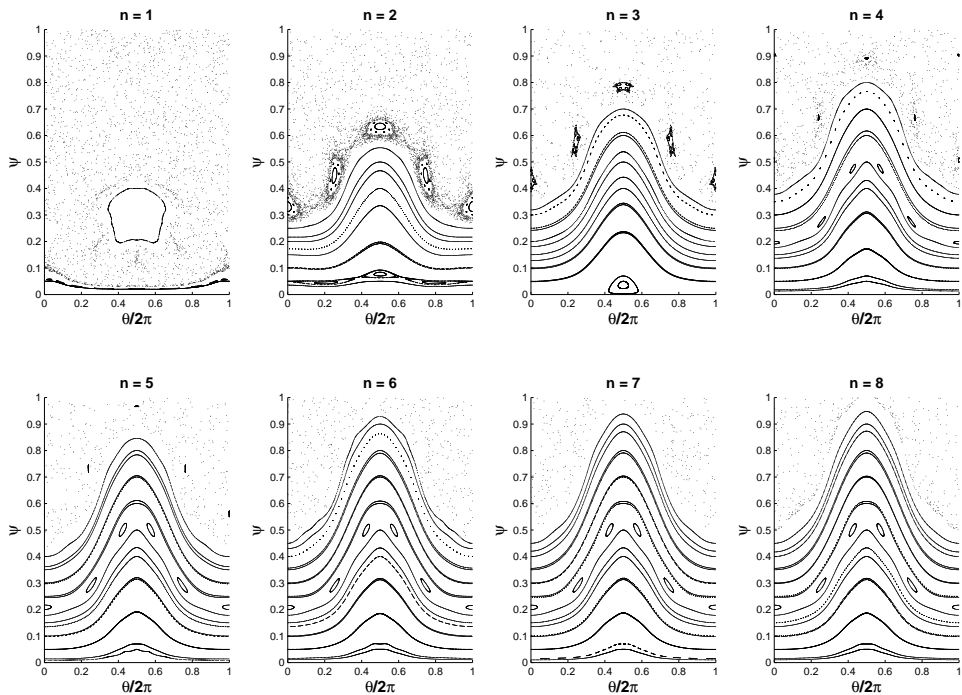}
    \caption{Phase space portraits of the symmetric tokamap (\ref{eq:tokamapSymmetricImplicit}) for different values of the order of degeneracy $n$ of the profile (\ref{eq:allureQPolynomiale}). Parameters are $K = 3$, $W_{0} = 5/2\pi$, $c  = 1.3$ and $\psi_0 = 0.5$.}
    \label{fig:influenceDegrePolynomeTokamapSymetrique}
\end{figure}

\begin{figure}[htbp]
    \centering
    \includegraphics[width = \textwidth]{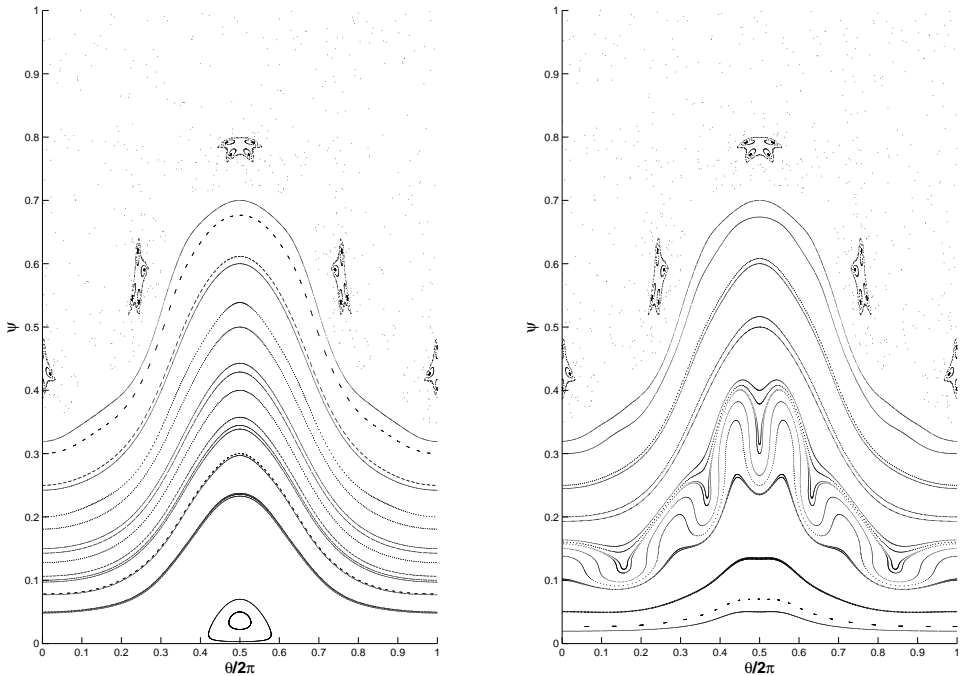}
    \caption{Phase space portraits of the symmetric tokamap (\ref{eq:tokamapSymmetricImplicit}). (left) $W_m(\psi)$ is given by (\ref{eq:allureQ3}); (right) $W_{rs}(\psi)$ is given by (\ref{eq:allureQ3Module}). Parameters $K = 3$, $W_{0} = 5/2\pi$, $c  = 1.3$ and $\psi_0 = 0.5$.}
    \label{fig:influenceMonotonieTokamapSymetrique}
\end{figure}

For every profile (\ref{eq:allureQPolynomiale}), with the degree $n$
going from 1 to 8, the phase portrait of the symmetric tokamap
(\ref{eq:tokamapSymmetricImplicit}) has been computed. The results
of this iterative process, for $\psi$ in the physical domain $[0,
1]$, are given for the same value $K=3$ of the stochasticity
parameter in Fig. \ref{fig:influenceDegrePolynomeTokamapSymetrique}.
Several comments are called for. First of all, it is clear from the
figure that the case $n=1$ is dramatically different from the
others. In the case $n=1$, the map is twist as $W'(\psi)$ does not
vanish. The figure exemplifies the \emph{dramatic improvement on
confinement induced by the non-twist condition} that is satisfied
when $n \geq 2$. This has been discussed recently on a general basis
by Rypina {\it{et al.}} \cite{RypinaRobustTransportBarriers2007}.
Secondly, when the non-twist condition is satisfied, the core
magnetic field lines are almost integrable. The central non-chaotic
region is all the broader in $\psi$-space as the $W$-profile is flat
around $\psi_0$, namely as $n$ is large. Yet this effect saturates
rapidly: for $n\geq6$, the phase space portraits are almost
indistinguishable. This is not surprising as the corresponding
$q$-profiles are indeed almost identical. Thirdly, there is a
perfect progression in the phase space appearances between the odd
and even $n$-exponents as $n$ increases in the non-twist case i.e.
for $n \geq 2$. This means that the fact that the $q$-profile has a
reversed shear or not is not important in terms of the confinement
of the magnetic field lines. What matters is the degree of
degeneracy of the zero of the shear profile or, so to speak, the
amount of local flatness of the $W$-profile. To check this
explicitly, let us consider the two following profiles
 \begin{equation}\label{eq:allureQ3}
 W_m(\psi) = -c (\psi - \psi_0)^3 + W_{0}
\end{equation}
\begin{equation}\label{eq:allureQ3Module}
 W_{rs}(\psi) = -c \vert(\psi - \psi_0)\vert^3 + W_{0}
\end{equation}
where $W_m$ (\ref{eq:allureQ3}) and $ W_{rs}$
(\ref{eq:allureQ3Module}) correspond respectively to a monotonic and
to a reversed-shear profiles with the same local flatness ($n=3$).
The results are displayed in figure
\ref{fig:influenceMonotonieTokamapSymetrique}.
This shows that the global behaviour of the magnetic field lines is
indeed the same in both cases: the outer stochastic region is
identical. There are only some limited apparent topological
discrepancies in the inner ordered zone. This third point is rather
novel since previous studies \cite{BalescuRevTokamap} commonly used
to focus solely on reversed-shear $q$-profiles to get the non-twist
condition.

\begin{figure}[htbp]
\begin{center}
\includegraphics[width= \textwidth]{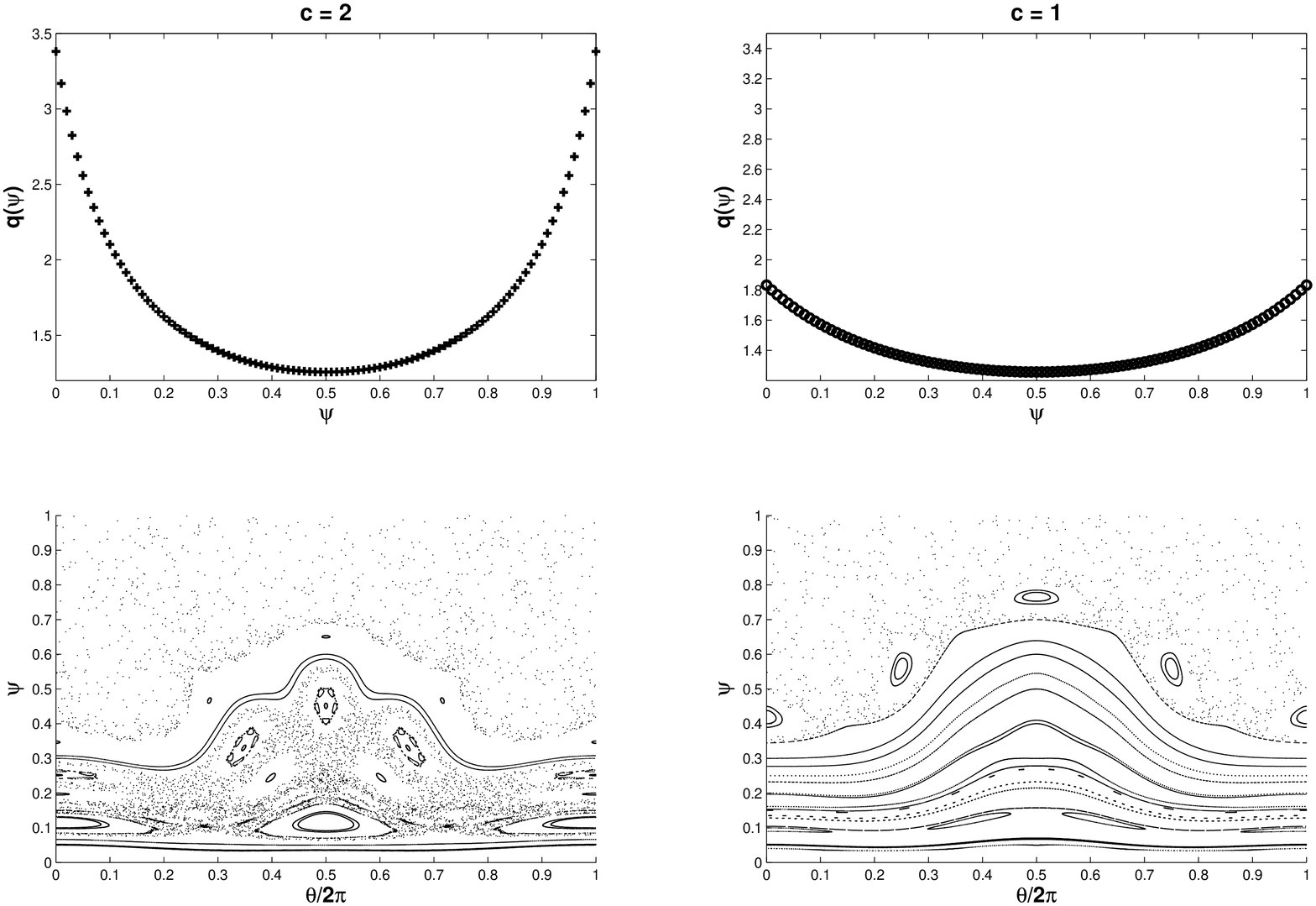}
\end{center}
\caption{Phase space portraits of the symmetric tokamap
(\ref{eq:tokamapSymmetricImplicit}) with
$W(\psi)=5/2\pi+c(\psi-.5)^{2}$ and $K=3$ for: (left) $c=2$, (right)
$c=1$.}
\label{fig:influenceDegrePolynomeQBordsIdentiqueTokamapSymetrique}
\end{figure}

\begin{figure}[htbp]
\begin{center}
\includegraphics[width= \textwidth]{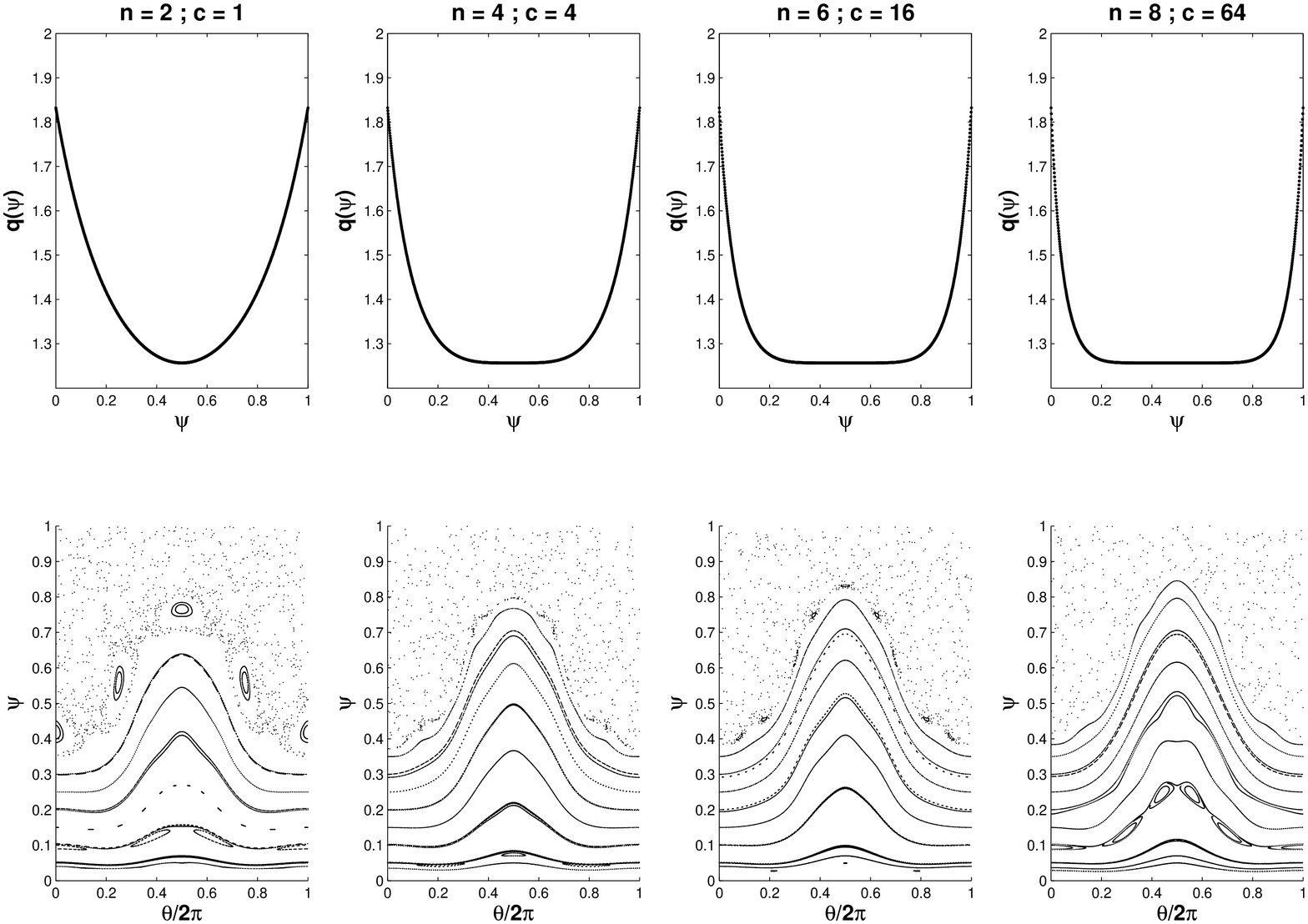}
\end{center}
\caption{(up) Reversed-shear $q$-profiles given by
$q(\psi)=1/[5/2\pi+c_{n}(\psi-.5)^{n}]$ with identical boundary
values and increasing central flatness (with $n$ even between 2 and
8) with (down) their corresponding phase space portraits of the
symmetric tokamap (\ref{eq:tokamapSymmetricImplicit}) for $K=3$.}
\label{fig:influenceFlatnessTokamapSymetrique}
\end{figure}

We have brought some numerical evidence of the influence on magnetic
confinement of the amount of flatness of the $q$-profile around its
extremum. In order to definitely ascertain this phenomenon, we
eventually wish to avoid the possible bias induced by using
different edge values of $W$ for each $n$ in Fig.
\ref{fig:influenceDegrePolynomeTokamapSymetrique}. It actually
happens that increasing the edge value of a polynomial $W$-profile
(\ref{eq:allureQPolynomiale}) with given degree $n$ may induce the
stochastization of the, otherwise regular, core magnetic field
lines. This is illustrated in Fig.
\ref{fig:influenceDegrePolynomeQBordsIdentiqueTokamapSymetrique} for
the case $n=2$.
Therefore, in order to single out the effect of the flatness of the
$W$-profile i.e. of the radial extent of the low-shear region around
an extremum, the study shown in Fig.
\ref{fig:influenceDegrePolynomeTokamapSymetrique} was revised for
the polynomial profiles (\ref{eq:allureQPolynomiale}) with variable
$c=c_{n}$ such that $W(0)=W(1)$ be a given constant for all $n$.
Results are depicted in Fig.
\ref{fig:influenceFlatnessTokamapSymetrique} for even values of $n$
between 2 and 8.
This figure shows that, \emph{all other things being equal}, the
radial extent of the regular magnetic field lines is all the wider
as the low shear region around the extremum of the $q$-profile is
broad.

\begin{figure}[htbp]
\begin{center}
\includegraphics[width= \textwidth]{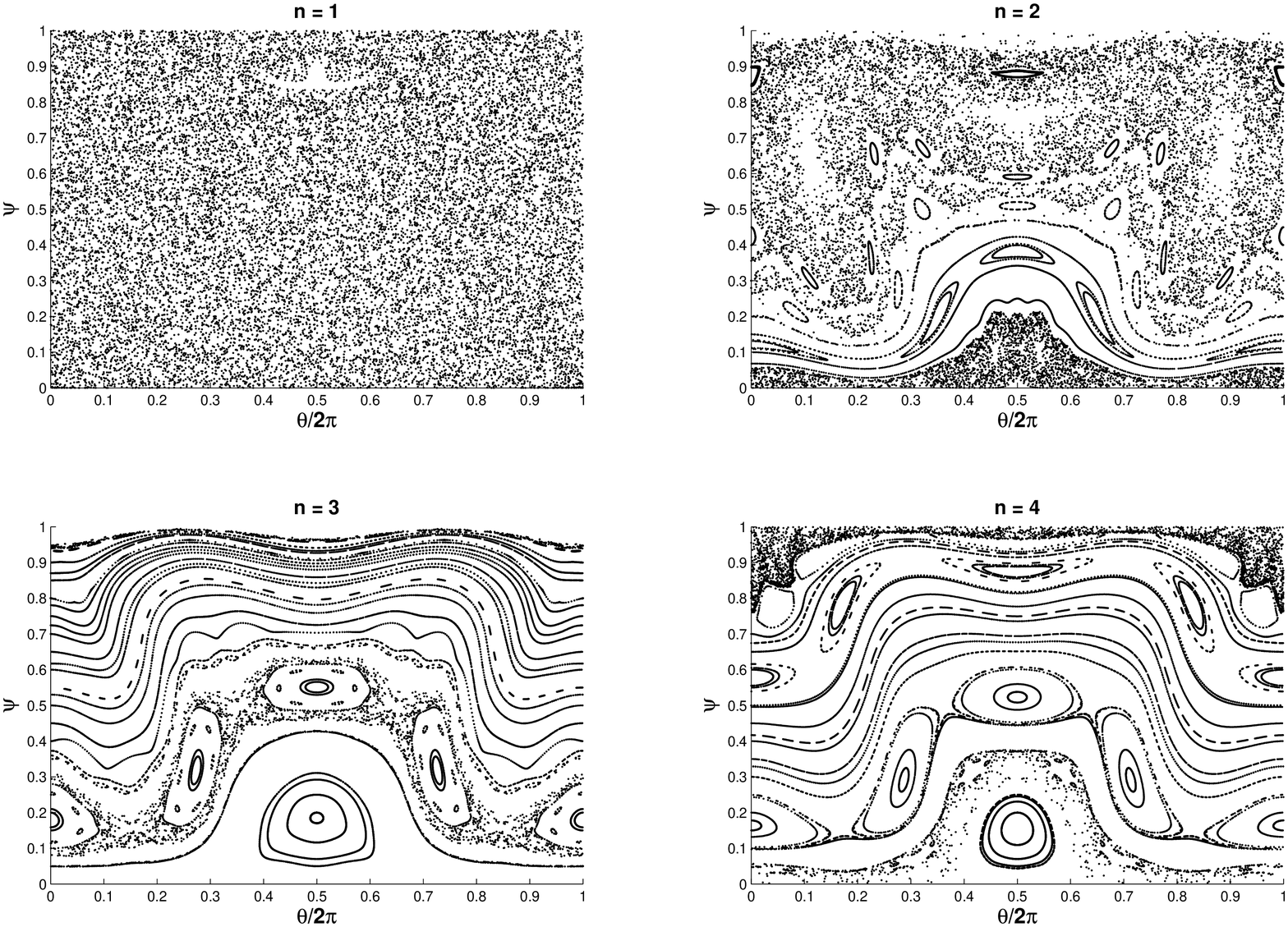}
\end{center}
\caption{Phase space portraits of the bounded symmetric tokamap
(\ref{eq:tokamapSymmetricBoundedImplicit}) for different values of
the order of degeneracy $n$ ($n=$ 1, 2, 3 and 4) of the profile
(\ref{eq:allureQPolynomiale}). Parameters are $K = 6$, $W_{0} =
5/2\pi$, $c  = 1.3$ and $\psi_0 = 0.5$.}
\label{fig:influenceFlatnessTokamapBoundedSymetrique}
\end{figure}

Finally, we wish to insist on the fact that the above conclusion
does not depend on the particular choice of the map. It remains
valid when considering the - less accurate - standard tokamap and
the bounded version of the symmetric tokamap. In this case, the
appearance of internal transport barriers can be seen more easily.
This can be shown in Fig.
\ref{fig:influenceFlatnessTokamapBoundedSymetrique}.
Whereas global chaos can be observed for $n=1$, the non-twist effect
that takes places for $n\geq2$ dramatically improves confinement. A
clear internal transport barrier emerges for $n=2$ separating the
internal and the outer chaotic regions. Increasing $n$, that is the
extent of the low shear region, induces a drastic reduction of
chaos.

\section{Influence of the radial extent of the low-shear region for some realistic reversed shear $q$-profiles}\label{sec:realisticProfile}

\begin{figure}[htbp]
    \centering
    \includegraphics[width = 0.8 \textwidth]{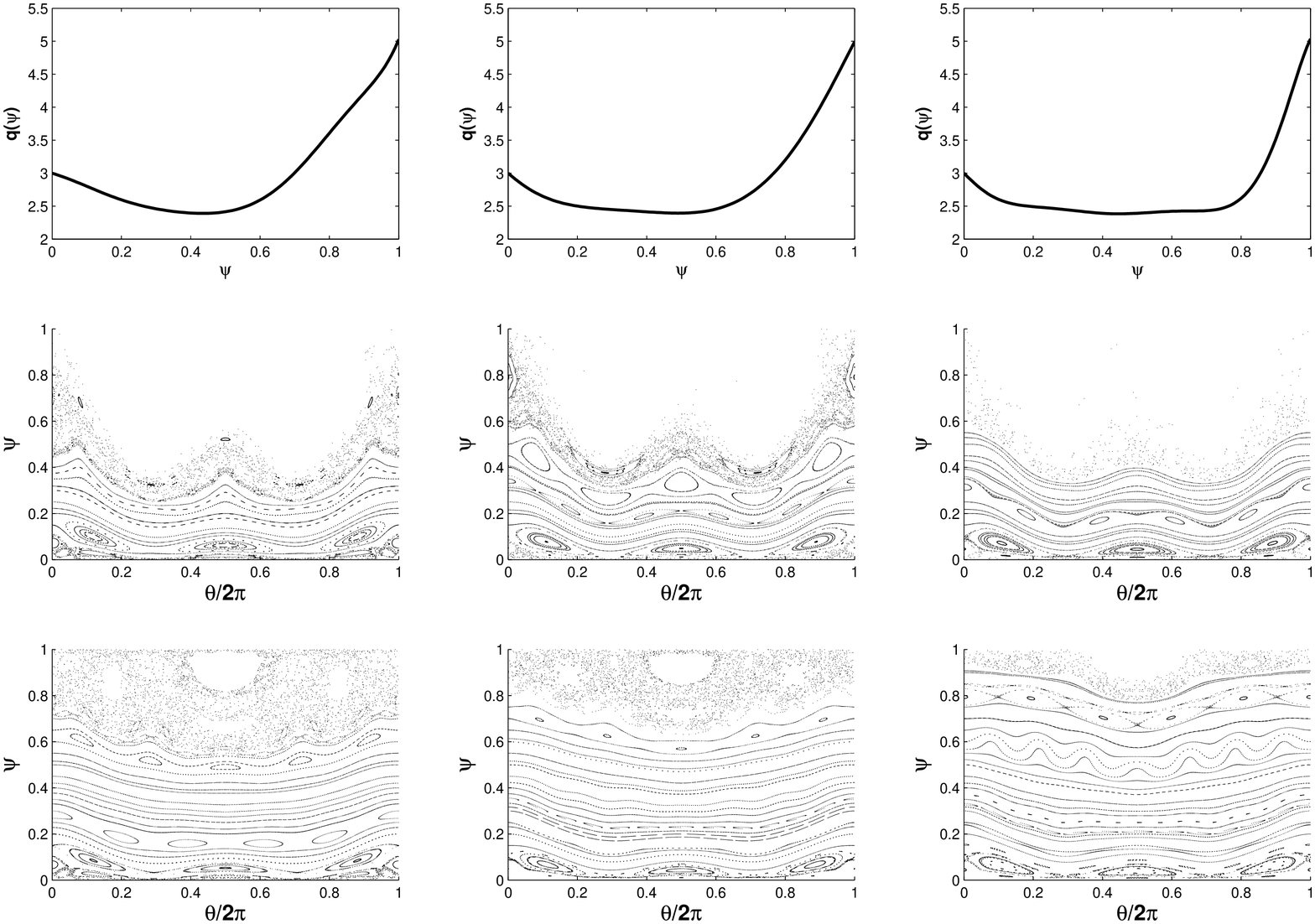}
    \caption{(top) $q$-profiles of increasing local flatness around $q_{min}$; (middle) corresponding phase space portraits of the symmetric tokamap (\ref{eq:tokamapSymmetricImplicit}) for 500 iterations and $K = 6$; (bottom) same as above for the bounded symmetric tokamap (\ref{eq:tokamapSymmetricBoundedImplicit}).}
    \label{fig:influencePlatitudeTokamapSymetrique}
\end{figure}

The previous academic case of polynomial centered $W$-profiles is
not realistic. We wish in this Section to briefly test our results
on the beneficial effect of the increased radial extent of the low
shear domain on more experimentally compatible $q$-profiles. We
shall here consider reversed shear profiles leading to ITBs. Typical
locations of the minimum of the $q$-profile correspond to a
normalized radius of 0.3-0.4 \cite{WolfRevue} and to a central $q(0)
\simeq 3$ while $q_{min}$, the minimum value of $q$, is about 2. We
have considered three $q$-profiles compatible with these realistic
data and computed the corresponding phase portraits for the magnetic
field lines. The results are given in Fig.
\ref{fig:influencePlatitudeTokamapSymetrique}. It is once again
clear from this figure that, for increasing flatness of $q$ about
its minimum and for constant $K$, the extent of the non-chaotic,
regular, magnetic field lines in the plasma core is broader.
In order to show that this effect is robust, namely map independent,
we have computed the phase space portraits for the symmetric bounded
tokamap (\ref{eq:tokamapSymmetricBoundedImplicit}) for the same
$q$-profiles. Results are plotted in Fig.
\ref{fig:influencePlatitudeTokamapSymetrique}. We do observe once
again that for increasing flatness there is chaos reduction.

A quantitative indicator of the degree of flatness of some
$W$-profile around its extremum located at $\psi _{0}$ could be
given by $f_{W}(\psi _{0})\equiv \lim_{\psi \rightarrow \psi
_{0}}\ln \left\vert W(\psi )-W(\psi _{0})\right\vert /\ln \left\vert
\psi -\psi _{0}\right\vert $. As shown previously on polynomial
profiles, for a given stochasticity parameter, confinement is
improved as $f_{W}(\psi _{0})$ increases. However, $f_{W}(\psi
_{0})$ gives the power of the first non-vanishing term in the Taylor
expansion of the $W$-profile around $\psi _{0}$. Since realistic
reversed-shear profiles are generically approximated locally by a
parabola, this would yield $f_{W}(\psi _{0})=2$. This has been
checked on the above profiles. It is thus necessary to produce a
practical indicator of the local amount of low shear. Let us then
assume that the reversed shear profile may be locally Taylor
expanded as
\begin{equation}
W(\psi )=W(\psi _{0})+\frac{1}{2}W^{\prime \prime }(\psi _{0})\delta
\psi ^{2}+O(\delta \psi ^{3}),  \label{Taylor_exp}
\end{equation}
with $\delta \psi \equiv \psi -\psi _{0}$. Then, for a given $W(\psi
_{0})$, the local flatness of $W$ increases as $\left\vert W^{\prime
\prime }(\psi _{0})\right\vert $ becomes smaller. This is in
agreement with the results given in Fig.
\ref{fig:influencePlatitudeTokamapSymetrique} where a diminution of
$W^{\prime \prime }(\psi _{0})$ corresponding to an increased local
flatness is observed from left to right. More precisely, one obtains
numerically $W^{\prime \prime }(\psi _{0})=10.6$ for the profile
located at the left of Fig.
\ref{fig:influencePlatitudeTokamapSymetrique}, $W^{\prime \prime
}(\psi _{0})=7.8$ for the middle profile and $W^{\prime \prime
}(\psi _{0})=7.4$ for the right one. Finally, let us note that in
this Figure as well as in Fig.
\ref{fig:influenceFlatnessTokamapBoundedSymetrique}, the
stochasticity parameter $K$, that is proportional to the relative
amount of magnetic perturbation \cite{nobleInternal}, has been taken
unrealistically large. This points to the fact that the effect of
low shear is much stronger than usual KAM results on the survival of
tori that would be obtained for monotone $q$-profiles and
corresponding twist maps [See e.g. the large scale chaos exhibited
in the tokamap phase portrait of Fig. 9 in the work by Misguich et
al. \cite{nobleInternal} that is obtained also for $K=6$].

\section{Conclusions} \label{sec:conclusions}

The Hamiltonian formalism has been used to account for the behavior
of magnetic field lines under some perturbations of given
equilibrium safety factor profiles. The corresponding symplectic
maps were the symmetric tokamap and its bounded version in which the
last magnetic surface cannot be crossed. In this time-independent
framework, we have shown the drastic improvement on the confinement
of magnetic field lines produced by the local vanishing of the shear
profile: the existence of some $\psi_0$ such that $q'(\psi_0)=0$ is
a non-twist condition that acts in the same way the safety profile
being monotonic or having reversed-shear. All other things being
equal, the beneficial effect of this vanishing of the shear profile
was shown to be increased if the radial extent of the low-shear
region was increased. To be specific, the low-shear region induces
the formation of a belt of robust KAM tori acting as an internal
transport barrier whose width is all the larger as the extent of the
low-shear region is broad. This conclusion has been derived here
without any consideration of dynamical stability. It is important to
note that there are additional dynamical arguments that favor a
reduction of (electrostatic) turbulence for low magnetic shear,
since it is well known that the linear growth rates of drift and
ballooning type modes vanish at zero shear \cite{Connor2004}.

Finally, we wish to conclude by discussing the validity of this
simplified purely magnetic approach. An obvious limitation of the
model comes from its lack of self-consistency. In particular, there
is no retroaction on the turbulence level. However, we would like to
stress that such a simplified picture, being in a Hamiltonian form
(due to the magnetic field being divergence-free), inherits from
robustness properties w.r.t. fluctuations. This is a general
property of Hamiltonian systems and ensures that their long-time
simulations are valid, at least in an average sense \cite{Stein}. In
the present model, this means that introducing some small
perturbation in the symplectic symmetric tokamap, such a
perturbation being e.g. a reduction of the magnetic turbulence
\cite{Zou95} or another perturbation that may reflect the effect of
electrostatic turbulence, the essential features of the model are
fully preserved.

It is now desirable to improve the connection of the present results
with tokamak experiments and complete the modeling to address
thermal transport. This would amount to couple the present magnetic
approach to some heat transport modeling. Nevertheless, we believe
that the benefits of the purely magnetic approach, initiated notably
by Balescu, have not been exhausted yet. The role of the value of
the $q$-profile in the low shear zone as well as the role of
edge-gradients of the $q$-profile are other challenging important
questions on which we shall be focusing our attention in a
shortcoming future.

\ack Fruitful discussions with M. B\'{e}coulet, X. Garbet, P.
Hennequin, H. L\"{u}tjens and J.-M. Rax are gratefully acknowledged.
The authors especially thank G. Steinbrecher for useful
communications on tokamap related works.

\clearpage

\bibliographystyle{unsrt}
\bibliography{bibliographieArticleITBQProfile}
\end{document}